\UseRawInputEncoding
\documentclass[aps,prl,groupedaddress,twocolumn,showpacs]{revtex4-1}
\usepackage[dvips]{graphicx}% Include figure files
\usepackage{dcolumn}% Align table columns on decimal point
\usepackage{bm}% bold math
\usepackage{amssymb}
\usepackage{epstopdf}
\usepackage{amsmath}
\usepackage{color}
\usepackage[dvipdfmx]{hyperref}
\hypersetup{
 setpagesize = false,
 colorlinks = true,
 citecolor = blue,
 linkcolor = blue,
 urlcolor = blue,
 }
\begin{document}
%\title{Two-dimensional cobalt carbonitride compounds CoN$_4$C$_{10}$, Co$_2$N$_8$C$_6$ and Co$_2$N$_6$C$_6$}
\title{Structural, electronic, magnetic properties of Cu-doped lead-apatite Pb$_{10-x}$Cu$_x$(PO$_4$)$_6$O }
\author{Jianfeng Zhang$^{1}$}
\author{Huazhen Li$^{2}$}
\author{Ming Yan$^{2}$}
\author{Miao Gao$^{3}$}
\author{Fengjie Ma$^{4}$}
%\email{gaomiao@nbu.edu.cn}
\author{Xun-Wang Yan$^{5}$}\email{yanxunwang@163.com}
\author{Z. Y. Xie$^{2}$}\email{qingtaoxie@ruc.edu.cn}
\date{\today}
\affiliation{$^{1}$Institute of Physics, Chinese Academy of Sciences, P.O. Box 603, Beijing 100190, China}
\affiliation{$^{2}$Department of Physics, Renmin University of China, Beijing 100872, China.}
\affiliation{$^{3}$Department of Physics, School of Physical Science and Technology, Ningbo University, Zhejiang 315211, China}
\affiliation{$^{4}$The Center for Advanced Quantum Studies and Department of Physics, Beijing Normal University, Beijing 100875, China}
\affiliation{$^{5}$College of Physics and Engineering, Qufu Normal University, Qufu, Shandong 273165, China}
\begin{abstract}
%The realization of superconductivity at room temperature and ambient pressure has always been a long-standing target in physics, chemistry, and materials science.
The recent report of superconductivity in the Cu-doped PbPO compound stimulates the extensive researches on its physical properties. Herein, the detailed atomic and electronic structures of this compound are investigated, which are the necessary information to explain the physical properties, including possible superconductivity. By the first-principles electronic structure calculations, we find that the partial replacement of Pb at $4f$ site by Cu atom, instead of Pb at $6h$ site, plays a crucial role in dominating the electronic state at Fermi energy. The $3d$ electronic orbitals of Cu atom emerge near the Fermi energy and exhibit strong spin-polarization, resulting in the local moment around the doped Cu atom. %The O with $\frac {1}{4}$ occupation in $4e$ site maintains the charge balance of parent Pb$_{10}$(PO$_{4}$)$_6$O.
Particularly, the ground state of Pb$_{10-x}$Cu$_x$(PO$_4$)$_6$O (x = 1) is determined to be a semiconducting phase, in good agreement with the experimental measurements.

\end{abstract}

%\pacs{74.70.Kn, 74.20.Pq, 61.66.Hq, 61.48.-c}

\maketitle
%% introduction of 2D materials, like 2016 PRB
%% manipulation the electronic structure, tailor elocate
%% magnetic 2D is rare
%% Here, magneism change from Cr - Mn Fe Ni

\section{Introduction}
Owing to the great scientific significance and giant application prospect, room-temperature superconductors have always been the long-standing dream in the fields of physics, chemistry, and materials science. In the past few years, some milestone achievements have been made in the superconductivity research area.
In 2015, Drozdov $et ~al.$ discovered the superconductivity at 203 K in the sulfur hydride system under high pressure \cite{Drozdov2015}.
In 2019, Somayazulu $et ~al$ reported the superconductivity above 260 K in Lanthanum superhydride LaH$_10$ at megabar pressures\cite{Somayazulu2019}. Subsequently, the discovery was confirmed by Drozdov $et ~al$ \cite{Drozdov2019}. In 2021, the superconductivity near room temperature in Yttrium Superhydride was found sucessively by Troyan $et ~al.$ \cite{Troyan2021}, Snider $et ~al.$\cite{Snider2021}, and Kong $et ~al.$\cite{Kong2021}.
The high-temperature superconductivity under high pressure environment is explained by the metalization of $\sigma$-bonding electronic states in the theoretical studies \cite{Miao2015}.
In the first half of 2023, the superconductivity in Lu-N-H system was reported by Dias' group \cite{Dasenbrock-Gammon2023}，but their results were questioned by other groups \cite{Ming2023}.
Apart from the hydrogen-rich superconductors, the superconductivity near 80 K in the nickelate La$_3$Ni$_2$O$_7$ under high pressure was found, which include $3d$ transition metal Ni and oxygen element\cite{Sun2023}. Very recently, Lee $et. al$ report the superconductor Pb$_{10-x}$Cu$_x$(PO$_4$)$_6$O (0.9 $\leq$ x $\leq$ 1.1)\cite{Lee2023,Lee2023a} and the critical temperature can reach up to 400 K. The results is so amazing and shocking that all people wonder whether and why the material has the high temperature superconductivity.
Because the crystal structure and detailed electronic structure are the key information to understand the formation of superconducting state,
 it is necessary and urgent to clarify these structural and electronic properties of Pb$_{10-x}$Cu$_x$(PO$_4$)$_6$O at present.

In the Pb$_{10}$(PO$_4$)$_6$O compound, there are two kinds of Pb atoms located in the $6h$ and $4f$ Wyckoff sites. In the first two papers\cite{Lee2023,Lee2023a}, the authors report that one of four Pb atom in the $4f$ site is replaced by Cu atom, and due to this substitution and the resulted distortion, the superconducting state occurs in the one-dimension chain of $4f$ Pb atom along $c$ axis. In the theoretical studies from Chen's group and Griffin\cite{lai2023,Griffin2023}, their first-principles electronic structure calculations are all based on the substitution of Pb at $4f$ site, and the Pb$_{10-x}$Cu$_x$(PO$_4$)$_6$O (x = 1) is metallic with the high density states at Fermi energy.
However,
%the detailed atomic structure of Cu-doped lead-apatite is still unclear, especially which kind of Pb atom is substituted by Cu atom. Our
our calculations indicate that the situation of the substitution of Pb atom at $6h$ site is more favorable in energy, and the ground state of Pb$_{10-x}$Cu$_x$(PO$_4$)$_6$O (x = 1) is semiconducting with the energy gap about 0.9 eV, which is in agreement with the experimental measurement in Ref. \citenum{Liu2023}.

\section{Computational details}
The plane wave pseudopotential method enclosed in Vienna Ab initio simulation package (VASP) \cite{PhysRevB.47.558, PhysRevB.54.11169} is used in our calculations. The generalized gradient approximation (GGA) with Perdew-Burke-Ernzerhof (PBE) formula \cite{PhysRevLett.77.3865} as well as the projector augmented-wave method (PAW) \cite{PhysRevB.50.17953} were adopted in ionic potentials.
The plane wave basis cutoff was 600 eV and the thresholds were 10$^{-5}$ eV and 0.01 eV/\AA ~ for total energy and force convergence.
 For structural optimizations, a mesh of $4\times 4\times 4$ k-points were sampled for the Brillouin zone integration, and the $8\times 8\times 12$ k-mesh for the self-consistent electronic structure calculations.
 %which was less than $2\pi \times $0.01/\AA~ in reciprocal space.
The electron correlation correction is taken into account in the GGA + U method with Hubbard U of 4.0 eV \cite{Cococcioni2005}.
%and also the nonempirical strongly constrained and appropriately normed (SCAN) meta-GGA method, GGA + U method are employed to consider the correction of electron correlation \cite{Sun2015,Cococcioni2005}(****).
\section{Results and discussion}

\begin{figure}
\begin{center}
\includegraphics[width=8.0cm]{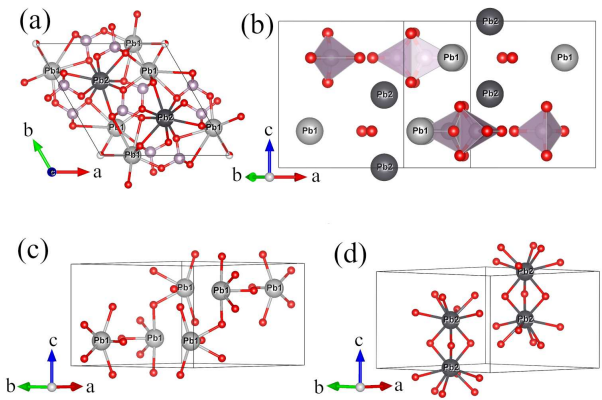}
\caption{Atomic structure of Pb$_{10}$(PO$_4$)$_6$O. (a) Top view, (b) Side view, (c) Pb(1) atom coordinated by five O atoms, (2) Pb(2) atom coordinated by nine O atoms.
 } \label{structmodel}
\end{center}
\end{figure}

The compound Pb$_{10}$(PO$_4$)$_6$O has a apatite-type structure, which crystalizes in the space group $P6_3/m$ (No. 176) with the crystal constants $a$ = 9.8650 \AA~ and $c$ = 7.4306 \AA~\cite{Krivovichev2003}. The detailed configurations are shown in Fig. \ref{structmodel}.
There two kinds of Pb atoms named Pb(1) and Pb(2) due to their inequivalent Wyckoff sites, which are located at the interstitial space among the PO$_4$ tetrahedrons.
We can view the Pb(1) atom and PO$_4$ tetrahedron as a layer in the $ab$ plane and these layers are stacked along $c$ axis direction, while Pb(2) atoms are sandwiched between two layers.
Pb(1) is bonded to five O atom in a 5-coordinate geometry. The lengths of these Pb-O bonds range from 2.44 - 2.63 \AA.  Pb(2) is bonded to nine O atoms in a 9-coordinate geometry. These Pb-O bond distances vary from 2.56 - 2.94 \AA.

\subsection{Which kind of Pb atom is replaced}
\begin{figure}
\begin{center}
\includegraphics[width=8.2cm]{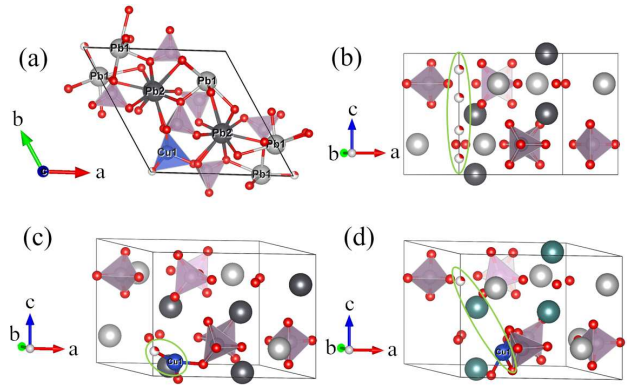}
\caption{(a) Positions of the substituted Cu atom in the optimized structure. (b) Four O$^{4e}$ positions in Pb$_{10}$(PO$_4$)$_6$O displayed in a green ellipse. (c) Positions of Cu and O$^{4e}$ atoms in the Cu$^{6h}$PbPO cell for O$^{4e}$(a) case, displayed in a green ellipse. (d) Positions of Cu and O$^{4e}$ atoms in the Cu$^{6h}$PbPO cell for O$^{4e}$(b) case, displayed in a green ellipse.
 } \label{fig2}
\end{center}
\end{figure}

For the convenience of calculations, we adopt $x$ = 1 in Pb$_{10-x}$Cu$_x$(PO$_4$)$_6$O to simulate the physical properties of the Pb$_{10-x}$Cu$_x$(PO$_4$)$_6$O (0.9 $\leq$ x $\leq$ 1.1) materials synthesized in experiments. There are two kinds of Pb atoms, Pb(1) at $6h$ site and Pb(2) at $4f$ site. The natural question is whether Pb(1) or Pb(2) is substituted by Cu atom.
With the measured and optimized crystal parameters, we carry out the energy calculations for the Cu-doped lead-apatite Pb$_{9}$Cu(PO$_4$)$_6$O.
The calculated energies are presented in Table. \ref{energy}, in which O$^{4e}(a)$ and O$^{4e}(b)$ represent the different positions of O atom in $4e$ site.
When substituting the Pb(1) at $6h$ site with Cu atom, rather than Pb(2) at $4f$ site, the Pb$_{9}$Cu(PO$_4$)$_6$O has the lower energy.
%When replacing Pb (1) at the 6h position instead of Pb2 at the 4f position, Pb9Cu (PO4) 6O has lower energy and is labeled as Cu6hPbPO.
In other words, the doped Cu atom should be located in the $6h$ site in the Cu-doped lead-apatite. Our findings is obviously different from the results from Lee $et~ al.$ and Griffin \cite{Lee2023, Lee2023a, Griffin2023}.
The Cu-doped lead-apatite with replacing Pb(1) and Pb(2) atoms are named as Cu$^{6h}$PbPO and Cu$^{4f}$PbPO, and the the structure of Cu$^{6h}$PbPO is displayed in Fig. \ref{fig2}(a).

\begin{table}
	\caption{The energies of the Cu$^{6h}$PbPO and Cu$^{4f}$PbPO compounds with the different positions of O atom are compared in the case of the optimized and fixed lattice parameters. The unit is eV/(formula cell). The energy values are from the PBE calculations.
	}
	\label{energy}
	\renewcommand\tabcolsep{11.5pt} % 调整表格列间的宽度
\begin{tabular*}{6.5cm}{ccc}
		\hline
		Fixed & O$^{4e}(a)$ & O$^{4e}(b)$   \\
		\hline
		Cu$^{6h}$PbPO   &-266.3510  & -266.1150    \\
		Cu$^{4f}$PbPO   &-265.7730 & -266.0160   \\
		\hline
\end{tabular*}
\begin{tabular*}{6.5cm}{ccc}
		\hline
		Optimized &  &    \\
		\hline
		Cu$^{6h}$PbPO   &-266.4862  & -266.4610  \\
		Cu$^{4f}$PbPO   &-265.7988  & -266.0167  \\
		\hline
\end{tabular*}

\end{table}

\subsection{ The detailed position of O$^{4e}$ atom}
The Wyckoff site $4e$ is occupied by $\frac{1}{4}$ O atom, which is related to four coordinates, (0.0 0.0 0.1342), (0.0 0.0 0.3658), (0.0 0.0 0.6342), and (0.0 0.0 0.8658), shown in Fig. \ref{fig2}(b). Because the substitution of Cu atom, the four $4e$ sites become inequivalent. It is necessary to examine the influence of O$^{4e}$ position on the electronic properties of Cu-doped PbPO compound. We select the coordinates (0.0 0.0 0.1342) and (0.0 0.0 0.8658) as the initial positions of O$^{4e}$ atom at the beginning of structural optimization. After optimization, the coordinates became (0.0 0.0 0.1991) and (0.0 0.0 0.7806) in the calculations, marked by O$^{4e}$(a) and O$^{4e}$(b) and shown in Fig. \ref{fig2}(c) and (d). From the Table. \ref{energy}, we can see that O$^{4e}$(a) configuration is more favorable energetically, which is related to the bonding between this O atom and the doped Cu atom.

\begin{figure}
\begin{center}
\includegraphics[width=8.0cm]{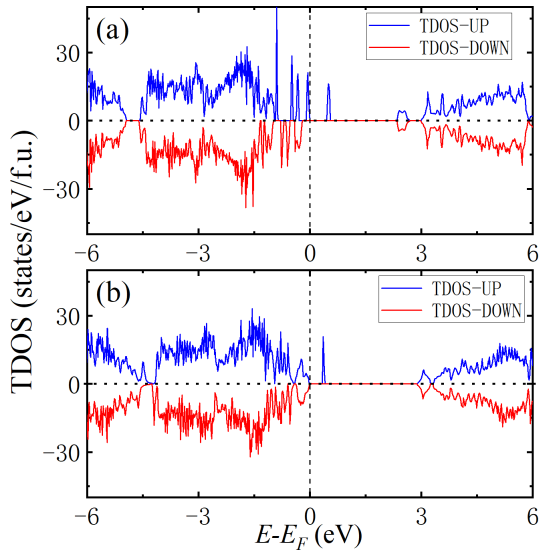}
\caption{Total density of states of Pb$_{9}$Cu(PO$_4$)$_6$O with the different positions of O atom from the standard PBE calculations, (a) O$^{4e}$(a), (b) O$^{4e}$(b).
 } \label{fig3}
\end{center}
\end{figure}

\subsection{Semiconducting ground state}
\subsubsection{PBE calculations}
Based on the above energy calculations, we determine that the Cu$^{6h}$PbPO with O$^{4e}$(a) occupation is the most possible structure of Pb$_{10-x}$Cu$_x$(PO$_4$)$_6$O (x = 1). Its total density of states are presented in Fig. \ref{fig3}(a), indicating that the Cu$^{6h}$PbPO with O$^{4e}$(a) occupation is semiconducting with the energy gap about 0.5 eV.
We also calculate the density of state for the Cu$^{6h}$PbPO with O$^{4e}$(b) occupation, shown in Fig. \ref{fig3}(b).  Although there are some obvious differences compared to the density of states spectra in Fig. \ref{fig3}(a), it is noticed that the Cu$^{6h}$PbPO with O$^{4e}$(b) occupation is also semiconducting with the energy gap about 0.5 eV. So far, we conclude that the Cu$^{6h}$PbPO systems has the semiconducting ground state in spite of the different O$^{4e}$ occupations.
\subsubsection{GGA + U calculations}
\begin{figure}
\begin{center}
\includegraphics[width=6.8cm]{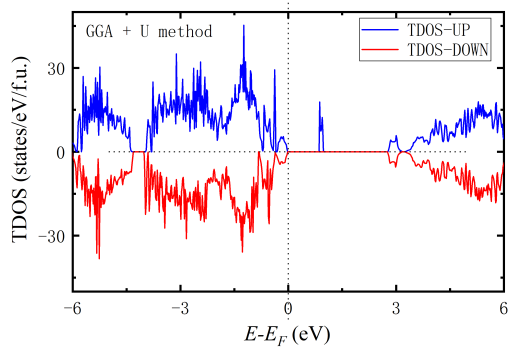}
\caption{Total density of states of Pb$_{9}$Cu(PO$_4$)$_6$O from GGA +U calculations with U = 4.0 eV.
 } \label{GGAU}
\end{center}
\end{figure}

To confirm the semiconducting ground state of Cu$^{6h}$PbPO with O$^{4e}$(a) occupation, the GGA + U method is employed to inspect its electronic structure. The Hubbard U value of 4.0 eV is adopted, which is an empirical value suggested in the previous studies \cite{Jain2011}. The total density of states are exhibited in Fig. \ref{GGAU}, in which the energy gap is about 0.9 eV, larger than the value of 0.5 eV derived from the PBE calculations.

%\subsubsection{SCAN calculations}

%\subsection{Influence of cell volume contraction on the electronic properties}
\subsection{Magnetic properties}
\begin{figure}
\begin{center}
\includegraphics[width=7.0cm]{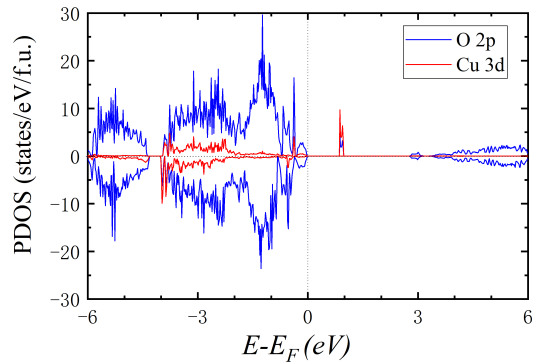}
\caption{Projected density of states on Cu $3d$ and O $2p$ orbitals from the GGA + U calculations.
 } \label{fig5}
\end{center}
\end{figure}

To exhibit the spin-polarized Cu $3d$ states and the hybridization of Cu $3d$ and O $2p$ orbitals, the total density of states are projected to atomic orbitals. Fig. \ref{fig5} shows the projected density of states on Cu $3d$ and O $2p$ orbitals. The density of states of O $2p$ orbitals are also presented to manifest that there is strong hybridization between O $2p$ and Cu $3d$ orbitals. The Cu $3d$ orbitals are obviously spin-polarized, which is related to the local moment of 1.0 $\mu_B$ around Cu atom.

To investigate the magnetic coupling between two doped Cu atoms, we construct the 1 $\times$ 1 $\times$ 2 and 2 $\times$ 1 $\times$ 1 magnetic supercells, in which the moments of two Cu atoms are aligned in the same or opposite direction. In the 1 $\times$ 1 $\times$ 2 supercell, two Cu atoms are in line with $c$ axis and their magnetic interaction represents the magnetic coupling along $c$ direction. Similarly, the interaction in a 2 $\times$ 1 $\times$ 1 supercell corresponds to the magnetic coupling in the $a$ axis direction (equal to $b$ axis). In our calculations, there is a weak ferromagnetic coupling along $c$ direction, and its energy is less than 0.1 meV. In the $ab$ plane, the coupling between two moments of Cu atoms is antiferromagnetic with the energy less than 1.0 meV. In a word, the magnetism of Cu-doped lead-apatite in the ground state is antiferromagnetic.

\section{Discussion and Conclusion}
Based on the first-principles calculations, the crystal, electronic, magnetic properties of Cu-doped lead-apatite are investigated. Our results indicate that in the Cu-doped lead-apatite Pb$_{10-x}$Cu$_x$(PO$_4$)$_6$O (x = 1), the doped Cu atom is located in the Wyckoff site $6h$ instead of the site $4f$. The electronic state near the Fermi energy is dominated by the Cu $3d$ and O $2p$ orbitals, and the spin polarization of Cu $3d$ electrons result in a local moment of about 1.0 $\mu_B$ around Cu atom. More importantly, the Pb$_{10-x}$Cu$_x$(PO$_4$)$_6$O (x = 1) is determined to have a semiconducting ground state with the energy gap of about 0.9 eV, in good agreement with the experimental measurements.

\begin{acknowledgments}
We sincerely thank Prof. Zhong-Yi Lu from Renmin University of China and Prof. Tao Xiang from the Institute of Physics in Chinese Academy of Sciences for their constructive discussions on the electronic states of the Cu-doped lead-apatite.
This work was supported by the National Natural Science Foundation of China under Grants Nos. 12274255, 12274458, 11974197, 11974207, 12074040 and the National R\&DProgram of China (Grants Nos. 2016YFA0300503, 2017YFA0302900).
\end{acknowledgments}

%\bibliography{ref-CrN4C2,Ref}
\bibliography{Ref}

\end{document}